\documentclass{amsart}%
\usepackage{amsfonts}%
\usepackage{amsmath}%
\usepackage{amssymb}%
\usepackage{graphicx}
\newtheorem{theorem}{Theorem}
\theoremstyle{plain}

\newtheorem{definition}{Definition}

\newtheorem{lemma}{Lemma}

\newtheorem{remark}{Remark}

\numberwithin{equation}{section}
\begin{document}
\title{New Scheme of Quantum Teleportation}
\author{A.Kossakowski\dag\ and M.Ohya\ddag}
\address{\dag Institue of Physics, N.Copernicus University, Torun, Poland\\
\ddag Department of Information Sciences, Tokyo University of Science Noda,
Chiba, Japan}
\maketitle

\begin{abstract}
A new scheme for quantum teleportation is presented, in which the complete
teleportation can be occurred even when an entangled state between Alice and
Bob is not maximal.

\end{abstract}

\section{Introduction}

The quantum teleportation has been discussed in several articles
\cite{Ben,AO,FO1,FO2,BK} by several different schemes. In most of models,
complete (perfect) teleportation can be occurred if the entangled state
between Alice and Bob is maximal. Here we reformulate the teleportation
process and show in our model that the complete teleportation is possible even
in the case for non-maximal entangled state.

\section{Basic Setting}

Let $\mathcal{H=}\mathbb{C}^{n}$ be a finite dimensional complex Hilbert
space, in which the scalar product
%TCIMACRO{\TEXTsymbol{<} }%
%BeginExpansion
$<$
%EndExpansion
,
%TCIMACRO{\TEXTsymbol{>} }%
%BeginExpansion
$>$
%EndExpansion
is defined as usual. Let $e_{n}$ ( $n=1$,$\cdots,n)$ be a fixed orthonormal
basis (ONB) in $\mathcal{H}$, and let $B(\mathcal{H)}$ be the set of all
bounded linear operators on $\mathcal{H}$, which is simply denoted by $M_{n}.
$In $M_{n}$, the scalar product ( , ) is defined by%

\[
(A,B):=trA^{\ast}B=\sum_{i=1}^{n}<Ae_{i},Be_{i}>
\]
Note that $e_{ij}:=\left\vert e_{i}\right\rangle \left\langle e_{j}\right\vert
$ $(i,j=1,\cdots,n)$ is a ONB in $M_{n}$ with respect to the above scalar
product. The mappings%

\[
A\in M_{n}\rightarrow A^{L}:=\sum\nolimits_{i=1}^{n}Ae_{i}\otimes e_{i}%
\in\mathcal{H\otimes H},
\]

\[
A\in M_{n}\rightarrow A^{R}:=\sum\nolimits_{i=1}^{n}e_{i}\otimes Ae_{i}%
\in\mathcal{H\otimes H}%
\]
define the inner product isomorphisms from $M_{n}$ into $\mathcal{H\otimes H}
$ such that%

\[
(A,B)=<<A^{L},B^{L}>>=<<A^{R},B^{R}>>,
\]
where the inner products in $\mathcal{H\otimes H}$ is denoted by
$<<\cdot,\cdot$ $>>$ .

Let $L(M_{n},M_{n})$ be the vector space of all linear maps $\Phi
:M_{n}\rightarrow M_{n}$. $M_{n}\otimes M_{n}$ is the set of all linear maps
from $\mathcal{H\otimes H}$ to $\mathcal{H\otimes H}$. By analogy between
$M_{n}$ and $\mathcal{H\otimes H}$, one can construct the inner product
isomorphisms between $L(M_{n},M_{n})$ and $M_{n}\otimes M_{n}$ such as%

\[
\Phi\in L(M_{n},M_{n})\rightarrow\Phi^{L}:=\sum\nolimits_{i,j=1}^{n}\Phi
e_{ij}\otimes e_{ij}\in M_{n}\otimes M_{n}%
\]

\[
\Phi\in L(M_{n},M_{n})\rightarrow\Phi^{R}:=\sum\nolimits_{i,j=1}^{n}%
e_{ij}\otimes\Phi e_{ij}\in M_{n}\otimes M_{n}%
\]

The inner products in $L(M_{n},M_{n})$ is defined as follows:%

\[
((\Phi,\Psi)):=tr\Phi^{\ast}\Psi=\sum\nolimits_{i,j=1}^{n}(\Phi e_{ij},\Psi
e_{ij}).
\]
One can easily verify that it is equal to%

\[
tr_{12}\Phi^{L\ast}\Psi^{L}=tr_{12}\Phi^{R\ast}\Psi^{R},
\]
where $tr_{12}$ is the trace over the space $M_{n}\otimes M_{n}$, whose ONB is
$\left\{  e_{ij}\otimes e_{kl}\right\}  .$

Let $\left\{  f_{\alpha};\alpha=1,\cdots,n^{2}\right\}  $ be another ONB in
$M_{n}$ so that one has $trf_{\alpha}^{\ast}f_{\beta}=\delta_{\alpha\beta}.$
It is easy to check that the maps $\Phi_{\alpha\beta}\in L(M_{n},M_{n})$
defined by $\Phi_{\alpha\beta}\left(  A\right)  :=f_{\alpha}Af_{\beta}^{\ast}$
for any $A\in M_{n}$ can be written as $\Phi_{\alpha\beta}=\left\vert
f_{\alpha}\right)  \left(  f_{\beta}\right\vert $ and the set $\left\{
\Phi_{\alpha\beta}\right\}  $ is a ONB of $M_{n}\otimes M_{n}. $ Moreover the
corresponding elements $\Phi_{\alpha\beta}^{L},\Phi_{\alpha\beta}^{R}\in
M_{n}\otimes M_{n}$ form ONBs of $M_{n}\otimes M_{n}. $The explicit expression
of $\Phi_{\alpha\beta}^{L}$ and $\Phi_{\alpha\beta}^{R}$ are%

\[
\Phi_{\alpha\beta}^{L}:=\sum\nolimits_{i,j=1}^{n}f_{\alpha}e_{ij}f_{\beta
}^{\ast}\otimes e_{ij}\text{ and }\Phi_{\alpha\beta}^{R}:=\sum
\nolimits_{i,j=1}^{n}e_{ij}\otimes f_{\alpha}e_{ij}f_{\beta}^{\ast}.
\]

There exist some important consequences for the above isomorphisms:

(1) Any map $\Phi\in L(M_{n},M_{n})$ is uniquely written as
\[
\Phi\left(  A\right)  =%
%TCIMACRO{\dsum }%
%BeginExpansion
{\displaystyle\sum}
%EndExpansion
c_{\alpha\beta}\Phi_{\alpha\beta}\left(  A\right)  =%
%TCIMACRO{\dsum }%
%BeginExpansion
{\displaystyle\sum}
%EndExpansion
c_{\alpha\beta}f_{\alpha}Af_{\beta}^{\ast}\text{ with some }c_{\alpha\beta}%
\in\mathbb{C}%
\]

(2) If $\Phi\left(  A^{\ast}\right)  =\Phi\left(  A\right)  ^{\ast},$ then
$c_{\alpha\beta}=\overline{c_{\alpha\beta}}$ $\in\mathbb{R}$ and $\Phi^{L}$,
$\Phi^{R}$ are self-adjoint in $\mathcal{H\otimes H}$.

(3) If $\Phi\left(  A\right)  =\Phi\left(  A\right)  ^{\ast},$that is, the
matrix $C:=$ $\left(  c_{\alpha\beta}\right)  $ is Hermitian, then $\Phi$ and
$\Phi^{L}$, $\Phi^{R}$ can be written in the following canonical forms:%

\[%
\begin{array}
[c]{c}%
\Phi\left(  A\right)  =%
%TCIMACRO{\dsum _{\alpha}}%
%BeginExpansion
{\displaystyle\sum_{\alpha}}
%EndExpansion
c_{\alpha}g_{\alpha}Ag_{\alpha}^{\ast}\\
\Phi^{L}=%
%TCIMACRO{\dsum _{\alpha,i,j}}%
%BeginExpansion
{\displaystyle\sum_{\alpha,i,j}}
%EndExpansion
c_{\alpha}g_{\alpha}e_{ij}g_{\alpha}^{\ast}\otimes e_{ij}\\
\Phi^{R}=%
%TCIMACRO{\dsum _{\alpha,i,j}}%
%BeginExpansion
{\displaystyle\sum_{\alpha,i,j}}
%EndExpansion
c_{\alpha}e_{ij}\otimes g_{\alpha}e_{ij}g_{\alpha}^{\ast}%
\end{array}
\]
where $\left\{  g_{\alpha};\alpha=1,\cdots,n^{2}\right\}  $ is some ONB in
$M_{n}$ and $c_{\alpha}\in\mathbb{R}$.

(4) From (3) it follows that for any ONB $\left\{  f_{\alpha}\right\}  $
\[
P_{\alpha}:=\Phi_{\alpha\alpha}^{L}=\sum\nolimits_{i,j=1}^{n}f_{\alpha}%
e_{ij}f_{\alpha}^{\ast}\otimes e_{ij},\text{ }Q_{\alpha}:=\Phi_{\alpha\alpha
}^{R}=\sum\nolimits_{i,j=1}^{n}e_{ij}\otimes f_{\alpha}e_{ij}f_{\alpha}^{\ast}%
\]
are mutual orthogonal projections in $\mathcal{H\otimes H}$ satisfying
\[
\sum\nolimits_{\alpha=1}^{n^{2}}P_{\alpha}=\sum\nolimits_{\alpha=1}^{n^{2}%
}Q_{\alpha}=I\otimes I\text{ }(I\text{ is unity of }M_{n})
\]

(5) A any state (density operator) $\sigma_{12}$ on $\mathcal{H\otimes H}$ can
be written in the form%

\[
\sigma_{12}=\sum\nolimits_{\alpha=1}^{n^{2}}\lambda_{\alpha}Q_{\alpha}%
=\sum\nolimits_{\alpha=1}^{n^{2}}\lambda_{\alpha}\sum\nolimits_{i,j=1}%
^{n}e_{ij}\otimes f_{\alpha}e_{ij}f_{\alpha}^{\ast}%
\]
with $\sum\nolimits_{\alpha=1}^{n^{2}}\lambda_{\alpha}=1$ and $\lambda
_{\alpha}\geq0.$ Put%

\[
\Theta\left(  A\right)  :=\sum\nolimits_{\alpha=1}^{n^{2}}\lambda_{\alpha
}f_{\alpha}Af_{\alpha}^{\ast}%
\]
for any $A\in M_{n}.$ Then $\Theta$ is a completely positive (CP) map on
$M_{n},$ and $\sigma_{12}$ is written as%

\[
\sigma_{12}=\sum\nolimits_{i,j=1}^{n}e_{ij}\otimes\Theta(e_{ij}).
\]

\bigskip

Let take $A\in M_{n}$ with $trA^{\ast}A=1,$ then $A^{L}$ ($A^{R})$ is a
normalized vector in $\mathcal{H\otimes H}$ and it defines a state $\sigma$ in
$\mathcal{H\otimes H}$ as $\sigma:=|A^{L}>><<A^{L}|.$

\begin{definition}
The above state $\sigma$ is maximal entangled if $A^{\ast}A=AA^{\ast}=\frac
{I}{n},$ equivalently, $A=\frac{1}{\sqrt{n}}U$ with some unitary operator $U$
in $\mathcal{H}$.
\end{definition}

\begin{remark}
One can construct an ONB $\left\{  f_{\alpha}=U_{\alpha}/\sqrt{n}%
;\alpha=1,\cdots,n^{2}\right\}  $ with unitary $U_{\alpha}.$Then the
corresponding projections $P$ and $Q$ given above (4) are maximal entangled states.
\end{remark}

\begin{definition}
The map $\Phi\in L(M_{n},M_{n})$ is said to be normalized if $\Phi\left(
I\right)  =I$ , base preserving if $tr\Phi\left(  A\right)  =trA$ for all
$A\in M_{n},$ selfadjoint if $\Phi\left(  A\right)  ^{\ast}=\Phi\left(
A^{\ast}\right)  $ for all $A\in M_{n},$ positive if $\Phi\left(  A^{\ast
}A\right)  \geq0$ for all $A\in M_{n}$ and completely positive if
$\sum\nolimits_{i,j=1}^{n}<x_{i},\Phi\left(  A_{i}^{\ast}A_{j}\right)  x_{j}>$
$\geq0$ for any $x_{i}$ $(i=1,\cdots,n)\in\mathcal{H}$ and any $A_{i}$
$(i=1,\cdots,n)\in M_{n}.$ Note that the canonical form of completely positive
map is given by $\Theta$ above.
\end{definition}

\section{Usual Scheme of Quantum Teleportation}

The quantum teleportation scheme is written as follows.

\begin{description}
\item[Step 0:] Alice has an unknown quantum state $\rho$ on an $N$%
--dimensional subspace a Hilbert space $\mathcal{H}_{1}$ and she was asked to
teleport it to Bob.

\item[Step 1:] For this purpose, we need two other Hilbert spaces
$\mathcal{H}_{2}$ and $\mathcal{H}_{3}$, $\mathcal{H}_{2}$ is attached to
Alice and $\mathcal{H}_{3}$ is attached to Bob. Take an entangled state
$\sigma$ on $\mathcal{H}_{2}\otimes\mathcal{H}_{3}$ having certain correlation
between Alice and Bob and prepare an ensemble of the combined system in the
state $\rho\otimes\sigma$ on $\mathcal{H}_{1}\otimes\mathcal{H}_{2}%
\otimes\mathcal{H}_{3}$.

\item[Step 2:] Alice performs a measurement of the observable $F:=\sum
z_{\alpha}P_{\alpha}$, involving only the $\mathcal{H}_{1}\otimes
\mathcal{H}_{2}$ part of the system in the state $\rho\otimes\sigma$. When
Alice obtains $z_{\alpha}$, according to the von Neumann (or Luder's) rule,
after Alice's measurement, the state becomes
\[
\rho_{\alpha}^{(123)}:=\frac{(P_{\alpha}\otimes\mathbf{1})\rho\otimes
\sigma(P_{\alpha}\otimes\mathbf{1})}{tr_{123}(P_{\alpha}\otimes\mathbf{1}%
)\rho\otimes\sigma(P_{\alpha}\otimes\mathbf{1})}%
\]
where $tr_{123}$ is the full trace on the Hilbert space $\mathcal{H}%
_{1}\otimes\mathcal{H}_{2}\otimes\mathcal{H}_{3}$.

\item[Step 3:] Bob is informed which outcome was obtained by Alice. This
information is transmitted from Alice to Bob without disturbance and by means
of classical tools.

\item[Step 4:] Having been informed an outcome of Alice's measurement, Bob
performs a corresponding unitary operation (key) onto his system. That is, if
the outcome was $z_{\alpha}$, Bob operates a unitary operator $W_{\alpha}$ and
change the state into
\[
W_{\alpha}^{\ast}\left(  tr_{12}\rho_{\alpha}^{(123)}\right)  W_{\alpha}.
\]
If this state is equal to the original state $\rho$ which Alice sent, then the
teleportation is succeeded.
\end{description}

Thus the problem of the quantum teleportation is that for any state $\rho$ in
$\mathcal{H}_{1}$ whether we can construct the entangled state between
$\mathcal{H}_{2}\ $and $\mathcal{H}_{3}$ and the key $\left\{  W_{\alpha
}\right\}  $ such that $W_{\alpha}^{\ast}\left(  tr_{12}\rho_{\alpha}%
^{(123)}\right)  W_{\alpha}=\rho.$

In some models \cite{Ben,AO,FO1} complete teleportation is possible if the
entangled state $\sigma$ used for the teleportation and the projection
$P_{\alpha}$ are maximally entangled.

\section{New Scheme of Entanglement and Teleportation}

We propose a new protocol for quantum teleportation. Let us take the
conditions that all three Hilbert spaces $\mathcal{H}_{1},\mathcal{H}%
_{2},\mathcal{H}_{3}$ are $\mathbb{C}^{n}.$Let the state $\sigma$ in
$\mathcal{H}_{2}\otimes\mathcal{H}_{3}=\mathbb{C}^{n}\otimes\mathbb{C}^{n}$ be%

\[
\sigma=\sum\nolimits_{i,j=1}^{n}e_{ij}\otimes\Theta(e_{ij})
\]
Here $e_{ij},\Theta$ are those given in Section 2 with an ONB\ $\left\{
f_{\alpha};\alpha=1,\cdots,n^{2}\right\}  $ but are defined on $\mathcal{H}%
_{2}$ and $\mathcal{H}_{3}$. We set an observable $F$ in $\mathcal{H}%
_{1}\otimes\mathcal{H}_{2}$ to be measured by Alice as follows:
\[
F=\sum_{\alpha}z_{\alpha}P_{\alpha}:=\sum_{\alpha}z_{\alpha}\sum
\nolimits_{i,j=1}^{n}g_{\alpha}e_{ij}g_{\alpha}^{\ast}\otimes e_{ij},
\]
where $\left\{  g_{\alpha};\alpha=1,\cdots,n^{2}\right\}  $ is another ONB of
$M_{n}.$ Then we define the teleportation map for an input state $\rho$ in
$\mathcal{H}_{1}$ and the measured value $z_{\alpha}$ of Alice by
\[
T_{\alpha}\left(  \rho\right)  :=tr_{12}(P_{\alpha}\otimes\mathbf{1}%
)\rho\otimes\sigma(P_{\alpha}\otimes\mathbf{1}).
\]

\begin{lemma}
The teleportation map $T_{\alpha}$ has the form $T_{\alpha}\left(
\rho\right)  =\Theta\left(  g_{\alpha}\rho g_{\alpha}^{\ast}\right)  $ for any
$\rho$ in $\mathcal{H}_{1}.$
\end{lemma}

\begin{proof}
One can write $T_{\alpha}\left(  \rho\right)  $ as

$T_{\alpha}\left(  \rho\right)  =\sum\nolimits_{i,j=1}^{n}\sum
\nolimits_{k,l=1}^{n}\sum\nolimits_{t,s=1}^{n}tr\left(  g_{\alpha}%
e_{ij}g_{\alpha}^{\ast}\rho g_{\alpha}e_{ts}g_{\alpha}^{\ast}\right)
tr\left(  e_{ij}e_{kl}e_{ts}\right)  \Theta\left(  e_{kl}\right)  $

$=\sum\nolimits_{i,j,s=1}^{n}tr\left(  g_{\alpha}e_{ij}g_{\alpha}^{\ast}\rho
g_{\alpha}e_{ti}g_{\alpha}^{\ast}\right)  \Theta\left(  e_{jt}\right)  $

$=\sum\nolimits_{i=1}^{n}<g_{\alpha}e_{i},g_{\alpha}e_{i}>\sum
\nolimits_{j,t=1}^{n}<e_{j},g_{\alpha}^{\ast}\rho g_{\alpha}e_{t}%
>\Theta\left(  e_{jt}\right)  $

$=\sum\nolimits_{j,s=1}^{n}tr\left(  g_{\alpha}^{\ast}\rho g_{\alpha}%
e_{jt}\right)  \Theta\left(  e_{jt}\right)  $

$=\Theta\left(  g_{\alpha}\rho g_{\alpha}^{\ast}\right)  $
\end{proof}

It is easily seen that $T_{\alpha}$ is completely positive but not trace
preserving. In order to consider the trace preserving map from $T_{\alpha},$
let us consider the dual map $\widetilde{T}_{\alpha}$ of $T_{\alpha}$, i.e.,
$trAT_{\alpha}\left(  \rho\right)  =:tr\widetilde{T_{\alpha}}\left(  A\right)
\rho.$ Indeed it is%

\[
\widetilde{T}_{\alpha}(A)=g_{\alpha}^{\ast}\widetilde{\Theta}\left(  A\right)
g_{\alpha},A\in M_{n}%
\]
where $\widetilde{\Theta}$ is the dual to $\Theta$;%

\[
\widetilde{\Theta}\left(  A\right)  =\sum\nolimits_{\alpha=1}^{n^{2}}%
\lambda_{\alpha}f_{\alpha}^{\ast}Af_{\alpha}.
\]
The map $\widetilde{T}_{\alpha}$ is normalizable iff $rank\widetilde
{T}_{\alpha}\left(  I\right)  =n,$ that is, the operator $\widetilde
{T}_{\alpha}\left(  I\right)  $ is invertible. Put%

\[
\kappa_{\alpha}:=\widetilde{T}_{\alpha}\left(  I\right)  .
\]
In this case the teleportation map $\widetilde{T}_{\alpha}$ is normalized as%

\[
\widetilde{\Upsilon}_{\alpha}:=\kappa_{\alpha}^{-\frac{1}{2}}\widetilde
{T}_{\alpha}\kappa_{\alpha}^{-\frac{1}{2}}.
\]
The dual map $\Upsilon_{\alpha}$ of $\widetilde{\Upsilon}_{\alpha}$ is trace
preserving and it has the form as%

\[
\Upsilon_{\alpha}\left(  \rho\right)  =\Theta\left(  g_{\alpha}\kappa_{\alpha
}^{-\frac{1}{2}}\rho\kappa_{\alpha}^{-\frac{1}{2}}g_{\alpha}^{\ast}\right)
=\sum\nolimits_{\beta=1}^{n^{2}}\lambda_{\beta}f_{\beta}g_{\alpha}%
\kappa_{\alpha}^{-\frac{1}{2}}\rho\kappa_{\alpha}^{-\frac{1}{2}}\left(
f_{\beta}g_{\alpha}\right)  ^{\ast}.
\]
It is important to note that this teleportation map is linear with respect to
all initial states $\rho.$

Let us consider a special case of $\sigma$ such that%

\[
\sigma=\sum\nolimits_{i,j=1}^{n}e_{ij}\otimes\Theta(e_{ij})\text{ with }%
\Theta\left(  \bullet\right)  :=f\bullet f^{\ast}\text{ and }trf^{\ast}f=1.
\]
That is, $\sigma$ is a pure state. In this case, one has%

\[
T_{\alpha}\left(  \rho\right)  =\left(  g_{\alpha}f\right)  \rho\left(
g_{\alpha}f\right)  ^{\ast}%
\]
and%

\[
\kappa_{\alpha}=\left(  g_{\alpha}f\right)  ^{\ast}\left(  g_{\alpha}f\right)
.
\]

\begin{remark}
If $g_{\alpha}=U_{\alpha}/\sqrt{n}$ and $f=V/\sqrt{n}$, where $U_{\alpha}$ and
$V$ are unitary operators, then $\kappa_{\alpha}=1/n^{2},$which corresponds to
the usual discussion.
\end{remark}

\bigskip Further, it follows that $\Upsilon_{\alpha}$ is trace preserving iff
$rank\left(  g_{\alpha}\right)  =rank\left(  f\right)  =n$, and in such a case
one has%

\[
\Upsilon_{\alpha}\left(  \rho\right)  =\left(  fg_{\alpha}\right)
\kappa_{\alpha}^{-\frac{1}{2}}\rho\kappa_{\alpha}^{-\frac{1}{2}}\left(
fg_{\alpha}\right)  ^{\ast}%
\]
Put
\[
W_{\alpha}:=fg_{\alpha}\kappa_{\alpha}^{-\frac{1}{2}},
\]
which is easily seen to be unitary. Thus we proved the following theorem.

\begin{theorem}
Given ONB $\left\{  g_{\alpha};\alpha=1,\cdots,n^{2}\right\}  $ and a vector
$f$ in $M_{n}$ on the n-dimensional Hilbert space, if $rank\left(  g_{\alpha
}\right)  =rank\left(  f\right)  =n$ is satisfied, then one can construct an
entangled state $\sigma$ and the set of keys $\left\{  W_{\alpha}\right\}  $
such that complete teleportation occurs.
\end{theorem}

Note here that our teleportation protocol is not required that the entangled
state is maximal for the complete teleportation. We will discuss an example of
this point in the next section.

\section{Example}

Let us construct an example as mentioned in Sec.3. That is, we construct an
entangled state given in the form: $\sigma=\sum\nolimits_{i,j=1}^{n}%
e_{ij}\otimes\Theta(e_{ij})$ with $\Theta\left(  \bullet\right)  :=f\bullet
f^{\ast}$ and $trf^{\ast}f=1.$Then it is possible in our protocol to teleport
completely by means of non-maximal entangled state $\sigma.$ The above state
$\sigma$ is pure, so that $\sigma$ is maximally entangled iff $f$ = $\frac
{U}{\sqrt{n}}$ with some unitary operator $u.$ Therefore if $rank\left(
f\right)  =n$ and $f\neq\frac{U}{\sqrt{n}},$ then $\sigma$ is not maximally entangled.

We will consider a bit more general question: For a ONB $\left\{  f_{\alpha
}\right\}  $ $(\alpha=1,\cdots,n^{2})$ in $M_{n},$whether can we construct
$n^{2}$ projections $Q_{\alpha}=\sum\nolimits_{i,j=1}^{n}e_{ij}\otimes
f_{\alpha}e_{ij}f_{\alpha}^{\ast}$ such that all $Q_{\alpha}$ are mutually
orthogonal and not maximally entangled. This question is reduced to find out
the basis $\left\{  f_{\alpha}\right\}  $ such that $rank\left(  f_{\alpha
}\right)  =n$ for any $\alpha$ and $f_{\alpha}\neq\frac{U}{\sqrt{n}}$ with
unitary $U.$

(1) A positive answer for the above question is given in the case $n=2,$ that
is, $M_{2}.$ Let $S_{\alpha}$ $(\alpha=0,1,2,3)$ are spin matrices;
\[
S_{0}=I,S_{1}=\left(  \
\begin{array}
[c]{cc}%
0 & 1\\
1 & 0
\end{array}
\right)  ,S_{2}=\left(  \
\begin{array}
[c]{cc}%
0 & -i\\
i & 0
\end{array}
\right)  ,S_{3}=\left(  \
\begin{array}
[c]{cc}%
1 & 0\\
0 & -1
\end{array}
\right)  ,
\]
and put%

\[
\omega_{\alpha}:=\frac{S_{\alpha}}{\sqrt{2}}(\alpha=0,1,2,3).
\]
Now we consider an orthogonal transformation $C:\mathbb{R}^{4}\rightarrow
\mathbb{R}^{4}.$ In terms of $C:=\left(  C_{\alpha\beta}\right)  $ one defines
a new basis $\left\{  f_{\alpha}\right\}  $ in $M_{2}$:%

\begin{equation}
f_{\alpha}=\sum\nolimits_{\beta=0}^{3}C_{\alpha\beta}\omega_{\beta}.
\end{equation}
Since $C_{\alpha\beta}$ is real and $\omega_{\alpha}=\omega_{\alpha}^{\ast},$
it implies that $f_{\alpha}=f_{\alpha}^{\ast}$ and the equality%

\[
\det f_{\alpha}=\frac{1}{2}\left(  C_{\alpha0}^{2}-\sum\nolimits_{\beta=1}%
^{3}C_{\alpha\beta}^{2}\right)  ,
\]
so that all $f_{\alpha}$ have rank 2 iff $\det f_{\alpha}\neq0.$ Such
$f_{\alpha}$ $(\alpha=0,1,2,3)$ generate the corresponding projection
$Q_{\alpha}=\sum\nolimits_{i,j=1}^{n}e_{ij}\otimes f_{\alpha}e_{ij}f_{\alpha
}^{\ast}$ on mutually orthogonal subspaces of $\mathbb{C}^{2}\otimes
\mathbb{C}^{2}$ such that $Q_{\alpha}$ $(\alpha=0,1,2,3)$ are non-maximal
entangled states iff the transformation $\left\{  \omega_{\alpha}\right\}  $
to $\left\{  f_{\alpha}\right\}  $ can not be generated by unitary $U$ such as
$U\omega_{\alpha}U^{\ast}=f_{\alpha}.$

From the orthogonality relation to $C,$ it follows that%

\[
C_{\alpha0}^{2}+\sum\nolimits_{\beta=1}^{3}C_{\alpha\beta}^{2}=1\text{ and
}\sum\nolimits_{\alpha=0}^{3}C_{\alpha0}^{2}=1.
\]
These relations tell us that $\det f_{\alpha}\neq0$ iff $C_{\alpha0}^{2}%
\neq\frac{1}{2}.$ Thus the relation $\sum\nolimits_{\alpha=0}^{3}C_{\alpha
0}^{2}=1$ implies that $\det f_{\alpha}\neq0$ iff $C_{\alpha0}^{2}>\frac{1}%
{2}.$

As an example, let us take the matrix $C$ as the form%

\[
C:=R_{01}\left(  \theta_{1}\right)  R_{02}\left(  \theta_{2}\right)
R_{03}\left(  \theta_{3}\right)  ,
\]
where $R_{ab}\left(  \theta\right)  $ is the rotation in $\left(  a,b\right)
$-plan with angle $\theta.$Then one finds%

\[
C=\left(
\begin{array}
[c]{cccc}%
c_{1}c_{2}c_{3} & -s_{1} & -c_{1}s_{2} & -c_{1}c_{2}s_{3}\\
s_{1}c_{2}c_{3} & c_{1} & -s_{1}s_{2} & -s_{1}c_{2}s_{3}\\
s_{2}c_{3} & 0 & c_{2} & -s_{2}s_{3}\\
s_{3} & 0 & 0 & c_{3}%
\end{array}
\right)  ,
\]
where $c_{i}:=\cos\theta_{i}$ and $s_{i}:=\sin\theta_{i}.$ It is easy to check
that $f_{\alpha}$ generate the projections $Q_{\alpha}$, whose corresponding
states are non-maximal entangled if $\left\vert s_{3}\right\vert >$ $\frac
{1}{2}.$ This inequality can be realized by taking $\theta_{3}$ properly,
e.g., $\frac{\pi}{6}<\theta_{3}<\frac{5\pi}{6}.$

(2) We can construct even simpler ONB $\left\{  f_{\alpha};\alpha
=0,1,2,3\right\}  $ generating non-maximal entangled states such as%

\begin{align*}
f_{0}  & =\left(  \
\begin{array}
[c]{cc}%
\cos\theta_{1} & 0\\
0 & \sin\theta_{1}%
\end{array}
\right)  ,f_{1}=\left(  \
\begin{array}
[c]{cc}%
-\sin\theta_{1} & 0\\
0 & \cos\theta_{1}%
\end{array}
\right)  ,\\
f_{2}  & =\left(  \
\begin{array}
[c]{cc}%
0 & \cos\theta_{2}\\
\sin\theta_{2} & 0
\end{array}
\right)  ,f_{3}=\left(  \
\begin{array}
[c]{cc}%
0 & -\sin\theta_{2}\\
\cos\theta_{2} & 0
\end{array}
\right)  .
\end{align*}
These are the rank=2 matrices for $0<\theta_{1},$ $\theta_{2}<\pi/2$, and they
generate a non-maximal entangled state when $\theta_{1},$ $\theta_{2}\neq
\pi/4.$

\end{document}